\documentclass[prb,twocolumn,english,showpacs]{revtex4-1}
\usepackage[pdftex]{graphicx}
\usepackage{amssymb}

\newcommand{\ket}[1]{|#1\rangle}

\newcommand{\sket}[1]{\ensuremath{\left|#1\right\rangle }}

\begin{document}

\title{Universal spin dynamics in two-dimensional Fermi gases}

\author{Marco Koschorreck$^1$, Daniel Pertot$^1$, Enrico Vogt$^1$, and Michael K{\"o}hl$^{1,2}$}


\affiliation{$^1$Cavendish\,Laboratory,\,University~of Cambridge, JJ Thomson Avenue, Cambridge CB30HE, United Kingdom\\$^2$ Physikalisches Institut, University of Bonn, Wegelerstrasse 8, 53115 Bonn, Germany}

%
\maketitle

\textbf{Harnessing spins as carriers for information  has emerged as an elegant extension to the transport of electrical charges\cite{Awschalom2007}. The coherence of such spin transport in spintronic circuits is determined by the lifetime of spin excitations and by spin diffusion. Fermionic quantum gases are a unique system to study the fundamentals of spin transport from first principles since interactions can be precisely tailored and the dynamics is on time scales which are directly observable\cite{Gensemer2001,Du2008,Du2009,Sommer2011,Sommer2011b,Bruun2011,Wulin2011,Ebling2011,Bruun2012,Enss2012}. In particular at unitarity, spin transport is dictated by diffusion and is expected to reach a universal, quantum-limited diffusivity on the order of $\hbar/m$. Here, we study the non-equilibrium dynamics of a two-dimensional Fermi gas following a quench into a metastable, transversely polarized spin state. Using the spin-echo technique\cite{Hahn1950}, we measure the yet lowest  transverse spin diffusion constant\cite{Leggett1968,Leggett1970} of $0.25(3)\,\hbar/m$. For weak interactions, we observe a coherent collective transverse spin-wave mode that exhibits mode softening when approaching the hydrodynamic regime.}

Studying transport in low-dimensional nanostructures has a long and rich history because of its non-trivial features and its relevance for electronic devices. The most common case, charge transport, has great technological implications and it determines the current-voltage characteristics of a device. With the development of the field of spintronics\cite{Awschalom2007}, however, also spin transport has moved into the focus of the research interest. Spin transport has unique properties, setting it aside from charge transport: Firstly, the transport of spin polarization is not protected by momentum conservation and is greatly affected by scattering\cite{Weber2005, Sommer2011}. Therefore, the question arises what is the limiting case of the spin transport coefficients when interactions reach the maximum value allowed by quantum mechanics? Secondly, unlike charge currents (which lead to charge separation and the buildup of an electrical field, counteracting the current), spin accumulation does not induce a counteracting force.
\begin{figure}
\includegraphics[width=.8\columnwidth]{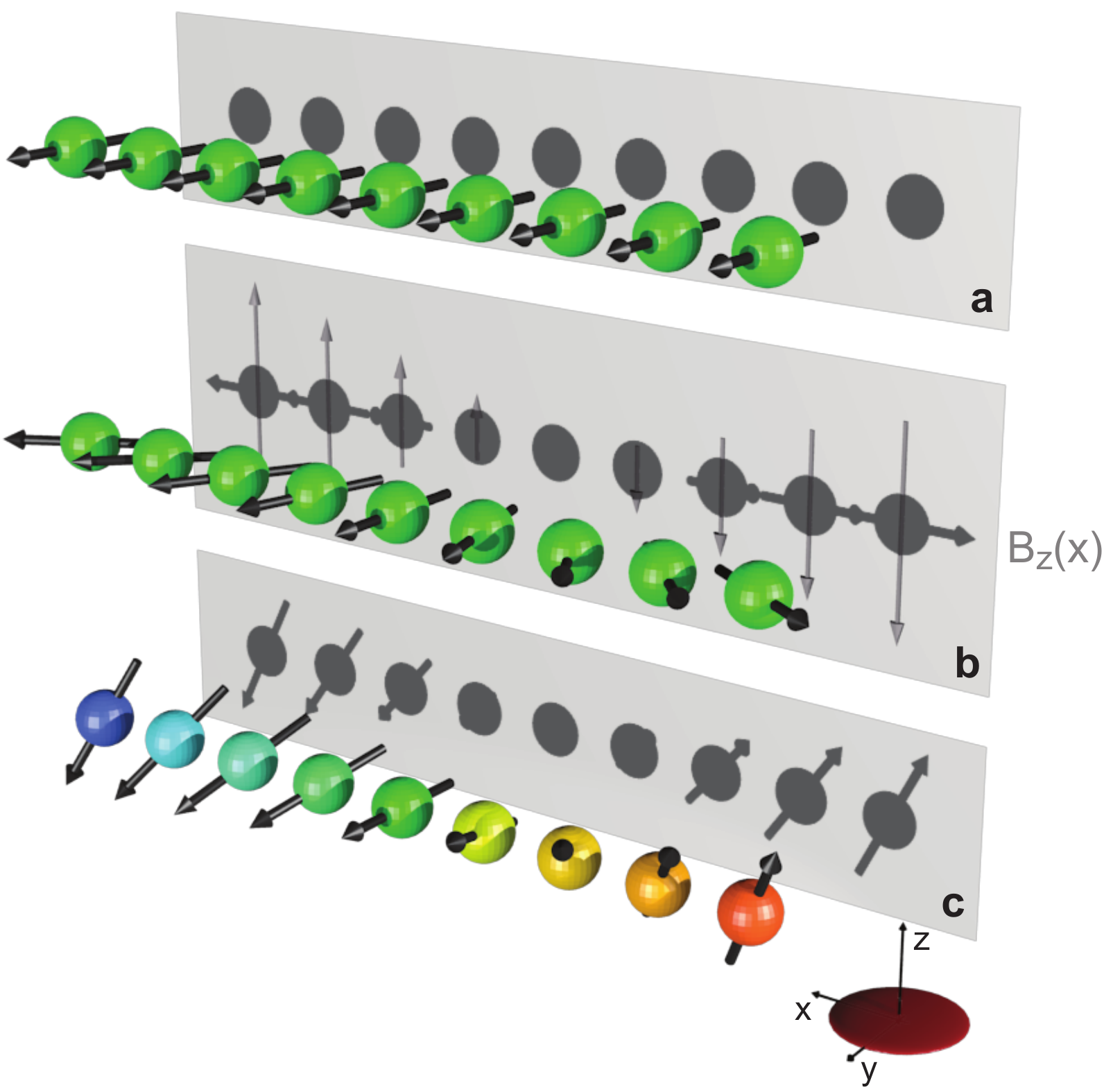}
\caption{Quench of a two-dimensional Fermi gas in which all atoms were initially prepared in the $\ket{\downarrow}$ state. (\textbf{a}), A $\pi/2$-pulse prepares  the Fermi gas polarized in the $S_y$-direction. (\textbf{b}) A magnetic field gradient $\partial B_z/\partial x$ causes the spins to acquire different phase-angles $\phi(x)$ in the equatorial plane. (\textbf{c}) Collisions cant the spins out of the equatorial plane due to the identical spin rotation effect. The acquired  projection along $S_z$ together with the motion of the atoms in the harmonic trap impedes rephasing of the spins when the magnetic field gradient is reversed. The spin states are illustrated in the rotating frame. }
\label{default}
\end{figure}

The main mechanism to even out a non-equilibrium magnetization $\vec{M}(\vec{r},t)= M(\vec{r},t)\,\vec{p}(\vec{r},t)$ is spin diffusion, which is an overall spin-conserving process. Other, spin non-conserving, processes are much slower in ultra cold Fermi gas experiments and are neglected hereafter. The gradient of the non-equilibrium magnetization $\nabla \vec{M}=\vec{p}\, \nabla M + M\nabla \vec{p}$ drives two distinct spin currents\cite{Dobbs2000}: The first term produces a longitudinal spin current parallel to $\vec{M}$ and the second term induces a transverse spin current orthogonal to $\vec{M}$. These spin currents are proportional to the longitudinal $\mathcal{D}_{\parallel}$ and transverse spin diffusivity $ \mathcal{D}_{\bot}$, respectively.

In general, the spin diffusivity in $D$ dimensions behaves as ${\mathcal{D}}=v\,l_\mathrm{mfp}/D$, where $v$ is the collision velocity and $l_\mathrm{mfp}$ is the mean free path. The mean-free path is determined by the elastic scattering cross section $\sigma$ between atoms. For short-range s-wave interactions at unitarity, the cross section attains its maximum value allowed by quantum mechanics $\sigma\sim 1/\lambda_{dB}^{D-1}$, where $\lambda_{dB}$ is the de-Broglie wavelength of the colliding particles. At high temperatures in the classical gas regime, the spin diffusion constant hence scales as ${\mathcal{D}} \propto T^{D/2}$ (up to logarithmic corrections in two dimensions owing to the energy dependence of the cross-section\cite{Bruun2012}). In the degenerate regime (of any dimensionality), the situation is quite different: the de-Broglie wavelength $\lambda_{dB}$ is of order $1/k_F$ and hence the mean-free path is $l_\mathrm{mfp}=1/(n\sigma) \approx 1/k_F$, where $k_F$ is the Fermi wave vector of the gas and $n\sim k_F^{D}$ is the density. Hence, the spin diffusion constant is given by Planck's constant $\hbar$ divided by the mass $m$ of the particle carrying the spin. This quantum limit can also be viewed as a result of the uncertainty principle by noticing that the mean-free path is limited by the mean interparticle spacing\cite{Enss2012}. The simple scaling argument, however, hides much of the rich underlying physics. In particular, it cannot explain the Leggett-Rice effect\cite{Leggett1968,Leggett1970,Corruccini1971}, the difference between longitudinal  and transverse  spin diffusivities\cite{Mullin1992}, and the transition to weak interactions where the physics changes because the system evolves from hydrodynamic to collisionless. So far, lowest spin diffusion constant has been measured for longitudinal spin currents to be $\mathcal{D}_{\parallel}=6.3\hbar/m$ in three-dimensional degenerate Fermi gases at unitarity\cite{Sommer2011}, approximately two orders of magnitude smaller than in semiconductor nanostructures\cite{Weber2005}.

Here, we study the coherence properties of a transversely polarized two-dimensional Fermi gas under the influence of magnetic field gradient. By employing the spin-echo technique \cite{Hahn1950} we gain access to the intriguing, but little understood, transverse spin dynamics in two-dimensional Fermi systems\cite{Dobbs2000}. The experiment starts with a polarised Fermi gas, where all spins are in the same state $\sket{\downarrow}$. Three resonant rf-pulses are applied to rotate the spin state by respective angles $\pi/2$ - $\pi$ - $\pi/2$ about a fixed axis in spin space, $S_x$. Consecutive pulses are separated by a time $\tau$ and we refer to the spin evolution time as $t=2\tau$. The initial $\pi/2$-pulse creates transverse spin polarisation  in a coherent superposition between \sket{\downarrow}  and \sket{\uparrow} (see Fig. 1a). An applied magnetic field gradient $B'\equiv\partial B_z/\partial x$ gives rise to a transverse spin wave (see Figure 1b) of wave vector $Q=\delta \gamma \, t B^\prime$, where  $\delta \gamma=152$\,kHz/G is the differential gyromagnetic ratio between $\sket{\uparrow}$ and $\sket{\downarrow}$ at 209\,G and $t$ is the evolution time (see Methods). On time scales much shorter than the trapping period, $Q$ is independent of the interaction strength\cite{Conduit2010}, and nearby atoms acquire a relative phase-angle of $\Delta\phi \approx Q/ k_F$. This lifts the spin-polarisation of the Fermi gas and the spins can collide with each other as they move in the harmonic potential or diffuse. Trivial dephasing induced by the magnetic field gradient is reversed by the $\pi$-pulse and the spin state refocus after a time $\tau$ if no decoherence has occured. The final $\pi/2$-pulse maps the spin state from the $S_y$ onto $S_z$, which is measured by performing a Stern-Gerlach experiment in time of flight, and we record $\langle M_z \rangle =(N_\uparrow-N_\downarrow)/(N_\uparrow+N_\downarrow)$.

Even though the interaction potential between atoms is not explicitly spin-dependent, an effective spin-exchange interaction is mediated by the required anti-symmetrisation of the scattering wave function. In binary collisions, this leads to the identical spin rotation effect \cite{Lhuillier1982}: in a collision both spins rotate about the axis defined by the sum of their spin orientations. In the Fermi degenerate regime, the binary collision picture has to be replaced by Landau's quasiparticle description. Here, quasiparticle excitations can be considered being affected by a ``molecular field'' resulting from the effective spin-exchange interaction\cite{Silin1958,Leggett1968,Leggett1970,Levy1984}. As a result, the spin wave cants out of the $S_x-S_y$--plane and forms a spin spiral which acquires a component along $S_z$ (see Figure 1c). The magnitude of the identical spin rotation effect is proportional to the mean-field interaction strength $g=-2 \pi \hbar^2/[m\ln(k_Fa_{2D})]$ of the gas. Here, $k_F$ is the Fermi wave vector for the initially spin-polarized sample, and $a_{2D}$ is the two-dimensional scattering length, which is linked to the binding energy of the confinement-induced dimer by\cite{Bloom1975} $E_B=\hbar^2/ma_{2D}^2$.

\begin{figure}
\includegraphics[width=.8\columnwidth]{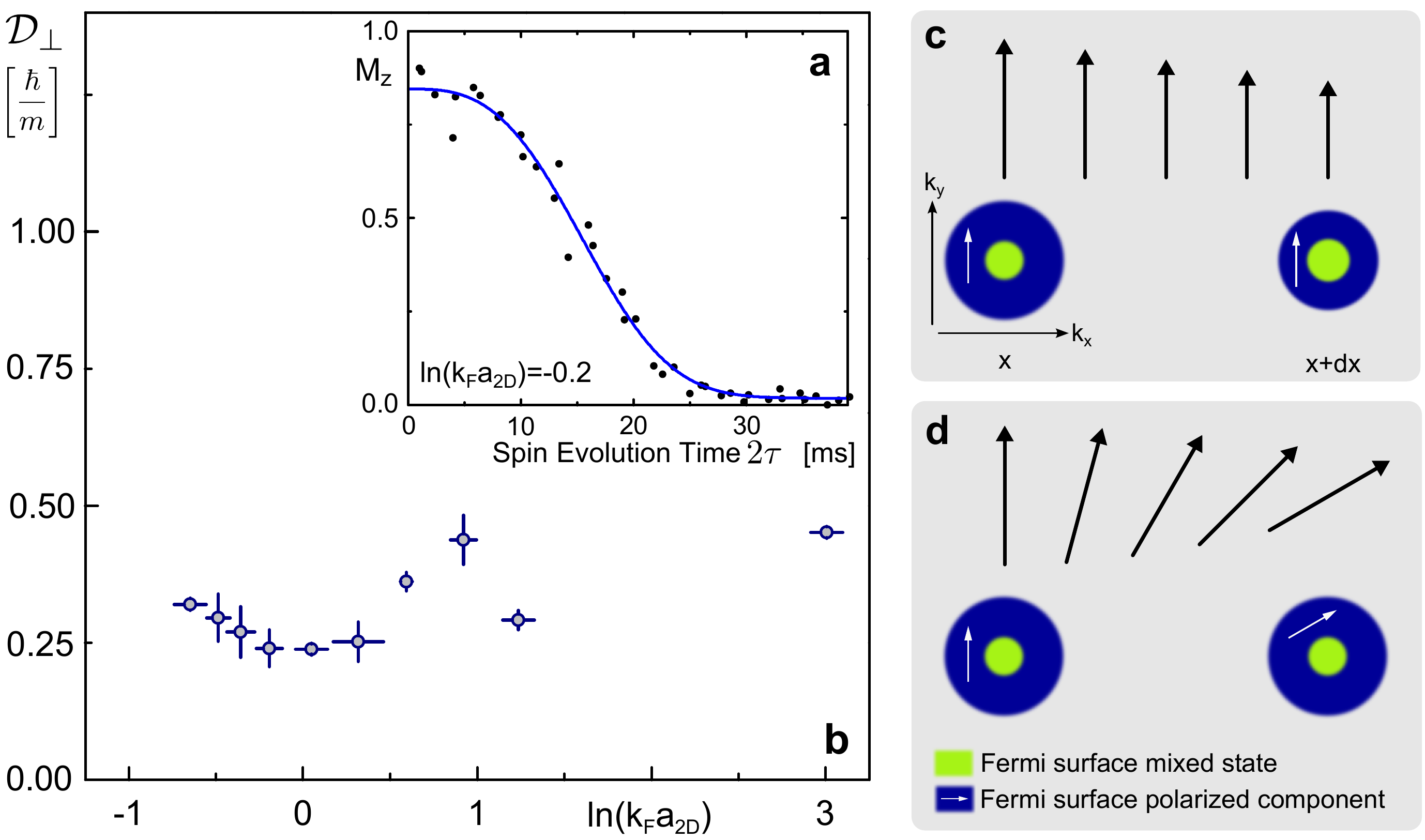}
\caption{Spin-echo signals in the strongly interacting regime. \textbf{a}, Spin echo-signal at $\ln(k_Fa_{2D})=-0.2$. The blue line is a fit $\propto \exp[-(2\Gamma \tau)^3]$. \textbf{b}, Transverse spin diffusion constant $\mathcal{D}_{\bot}$ as a function of interaction strength deduced from the decay constant $\Gamma$ of the spin-echo signal. Illustration of the different spatial variation of Fermi surfaces of a polarized Fermi gas for the case of longitudinal (\textbf{c}) and transverse (\textbf{d}) spin diffusion.}\label{default}
\end{figure}

In the strongly interacting regime, i.e., $-1<\ln(k_Fa_{2D})<3$, spin transport is dominated by diffusion and the spin-echo signal attains a  characteristic non-exponential decay of the form\cite{Hahn1950,Leggett1968,Leggett1970} $\propto \exp[-(2\Gamma \tau)^3]$. Fitting the envelope of the spin-echo signal\footnote{Our experimental timing is not phase stable with respect to the	Larmor precession of 50 MHz. Our analysis hence focusses
on the envelope of the spin-echo from which the shown data points typically scatter less than 10 percent. The combined error of preparation and detection of the magnetization for a single data point is less than 1 percent} (see Figure 2a) with this function we deduce the transverse spin diffusion constant from the measured decay rate as\cite{Hahn1950,Carr1954} ${\mathcal{D}}_{\bot}=12 \hbar^2\Gamma^3/(\delta\gamma\,B^\prime)^2$. We extract the transverse spin diffusion constant for various interaction strengths and find ${\mathcal{D}}_{\bot} = 0.25(3) \hbar/m$ at a shallow minimum around $\ln(k_Fa_{2D})=0$ (see Figure 2b). From the arguments given at the beginning, we expect the diffusivity to be on the order of $\hbar/m$. Observing a smaller value than for longitudinal spin diffusion in three dimensions\cite{Sommer2011} is probably less linked to the dimensionality rather than to the phase space available for collisions necessary to drive spin diffusion\cite{Mullin1992}: In the case of longitudinal spin currents, a gradient of the $M_z(x)$ polarization along the x-direction can be considered as a spatial variation of the local Fermi surfaces $k_{F,\uparrow}(x)$ and  $k_{F,\downarrow}(x)$ (see Fig. 2c). Only $\sket{\uparrow}$ spins in a small region near the Fermi surface can diffuse from $x$ to $ x+dx $, invoking the typical $T^{-2}$ scaling for quasiparticles in the deeply degenerate regime. In contrast, in the case of transverse spin currents the Fermi surfaces at different positions are of the same size but have slightly different directions of magnetization. (see Fig. 2d). Hence, a spin moving along $x$ from anywhere between the Fermi surfaces for spin-up and down must scatter to reach local equilibrium, which scales as $n_{\uparrow}-n_{\downarrow}$ and provides a much larger phase space. The result is that for a degenerate system the transverse diffusivity is smaller than the longitudinal and becomes independent of temperature\cite{Mullin1992}.

\begin{figure}
\includegraphics[width=0.95\columnwidth]{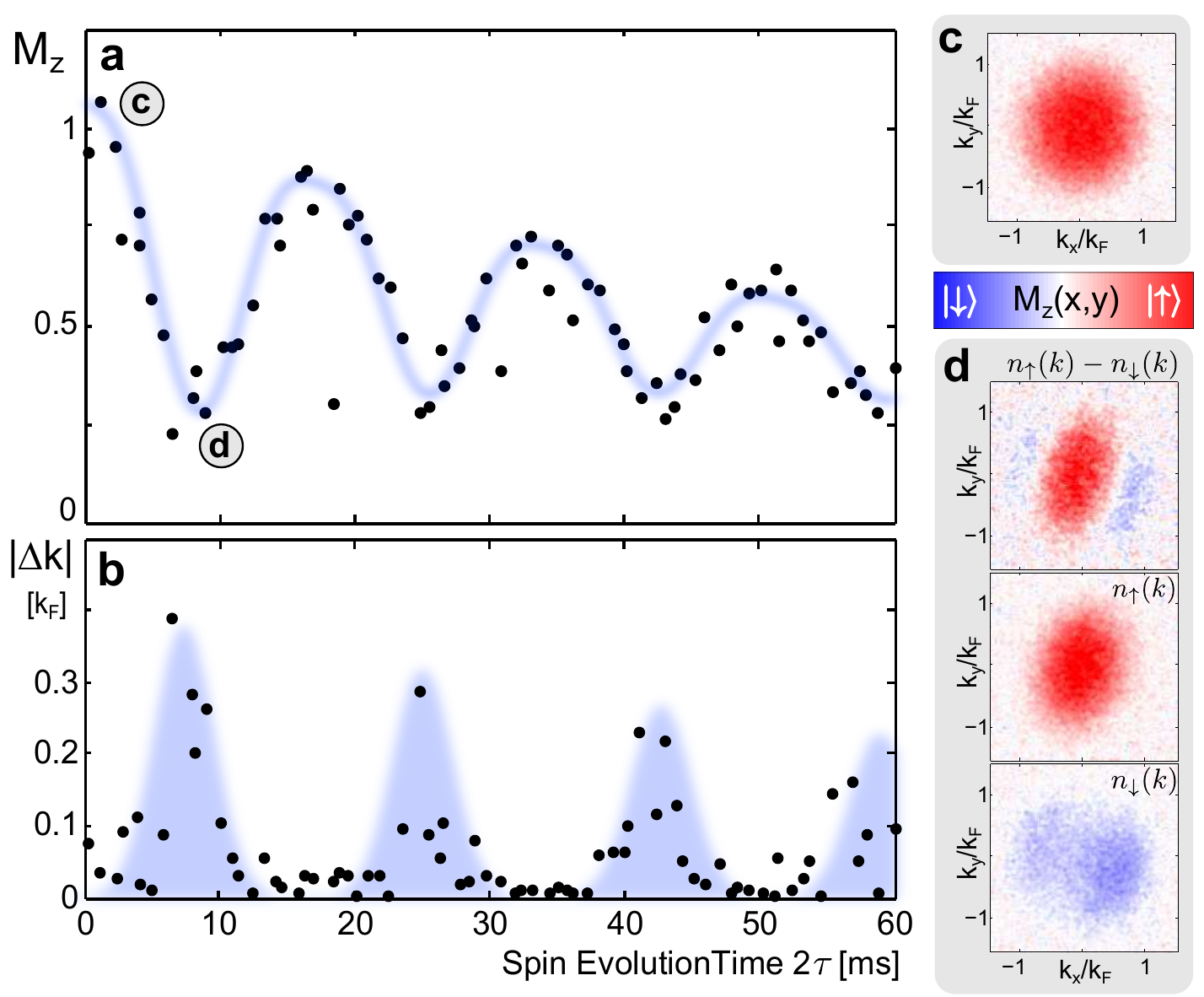}
\caption{ Observation of spin wave for the non-interacting gas, i.e., $\ln(k_Fa_{2D})\rightarrow+\infty$: \textbf{a}, We plot the envelope of the spin-echo signal (dots) and show the fit function (grey line). \textbf{b}, centre of mass separation of $\sket{\uparrow}$ and $\sket{\downarrow}$ after $11.8\,$ms of time of flight. \textbf{c} and \textbf{d}, momentum distribution of local spin polarisation for large and small spin polarisation, respectively. The lower two panels in \textbf{d} show the distribution of spin up and down, respectively. }
\label{default}
\end{figure}

For a repulsively interacting Fermi gas in three dimensions the decay of a spin-spiral state into a Stoner itinerant ferromagnet was considered\cite{Conduit2010}. In this case, the growth rate of the instability towards the formation of the ferromagnet order was found to scale $\propto Q^2$ and hence the spin-polarisation signal would decay $\propto \exp[-(\alpha t)^3]$, similar to spin diffusion.

In the weakly interacting attractive Fermi liquid regime\cite{Frohlich2012}, i.e., $\ln(k_Fa_{2D})>3$, we observe a qualitatively different behaviour (see Figure 3a). The envelope of the spin-echo signal decays exponentially and is modulated by a slow oscillation and both the frequency and the damping constant of the oscillation depend on the interaction strength. We attribute this slow periodic modulation to transverse spin waves which are excited by the spin-echo sequence. The spin current induced by the magnetic field gradient can only be inverted completely by the $\pi-$pulse when applied in phase with the harmonic motion. This can be qualitatively (for zero interactions) understood in a phase-space picture by considering that the magnetic field gradient causes a spatial offset between the potentials for the $\sket{\uparrow}$ and the $\sket{\downarrow}$ spin states and, correspondingly, the phase-space trajectories of the two spin-components are displaced\cite{Ebling2011}. Indeed, we observe that the difference of the center-of-mass momenta $\langle k_\uparrow\rangle-\langle k_\downarrow \rangle$ of the two spin densities $n_{\uparrow}(k)$ and $n_\downarrow(k)$ oscillates out-of-phase with the contrast of the spin-echo signal (Figure 3a and 3e). The total density profile $n(k)=n_{\uparrow}(k)+n_\downarrow(k)$ is stationary, which indicates a pure spin mode.

The measured envelope of the spin-echo signal (see Fig. 3a) is fitted with an empirical function of the form:  $A\exp(-\gamma t) \left(1-B|\sin(\omega t/2)|^3 \right)+C$. Here, $|A|,|B| \sim \mathcal{O}$$(1)$ and $|C| \ll1$ are amplitudes and a global offset of the spin echo signal, respectively. We find that this periodic function of frequency $\omega$ is approximating the non-sinusoidal signal best. The exponent is fixed to the value 3 for all interactions, and we do not see a large influence on the deduced parameters. Both the frequency $\omega$ and the exponential damping $\gamma$  depend on the interaction strength of the gas.

\begin{figure}
\includegraphics[width=0.8\columnwidth]{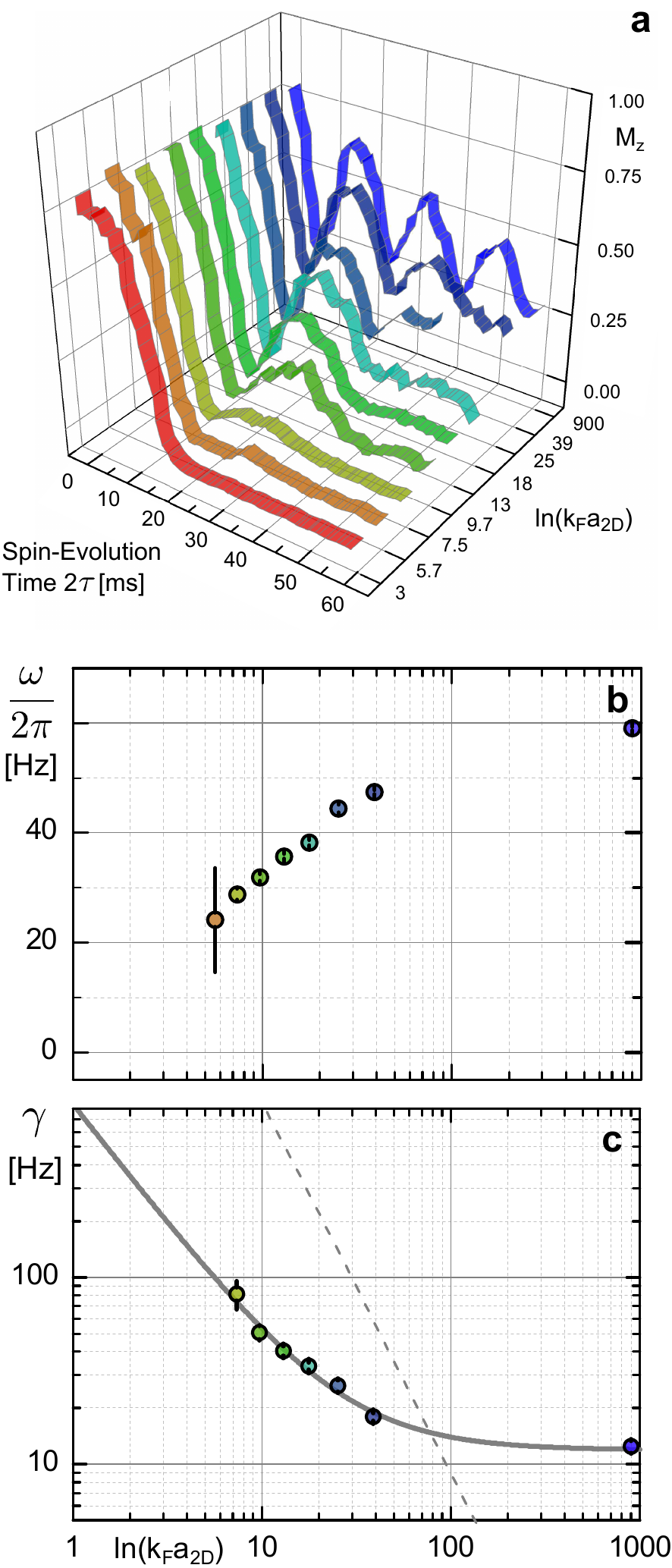}
\caption{Spin wave in the collisionless regime as function of interaction strength. \textbf{a}, Time evolution of global spin polarization for different interaction strengths. \textbf{b}, Frequency of oscillating spin-polarization. For the data point at $\ln(k_Fa_{2D}=3)$ we were not able to fit a frequency and set the value to zero.   \textbf{c} Damping coefficient $\gamma$ of the spin-polarization. The dashed line indicates the elastic scattering rate and the solid line is a fit to the measured decay.}
\label{default}
\end{figure}

For zero interaction strength the mode is at half the trap frequency and we observe a softening of the spin mode when approaching the hydrodynamic regime (Figure 4a and b). Such a behaviour is reminiscent of the collisionless-to-hydrodynamic transition of the spin-dipole oscillation\cite{Vichi1999,Sommer2011}, whereas spin-symmetric modes have non-zero frequencies even in the hydrodynamic regime\cite{Vogt2012}. The decay rate $\gamma$ of the oscillation (see Figure 4c) is found smallest for the non-interacting gas, where it is two orders of magnitude smaller than the dephasing rate $1/\tau_\mathrm{Ramsey}$ in the inhomogeneous magnetic field  (see Methods). As the interaction strength is increased, the rate $\gamma$ is growing (see Fig. 4c). We fit this growth with a power-law behaviour $\gamma= \alpha/\ln(k_Fa_{2D})^\beta+\gamma_0$ and we determine the exponent to be $\beta=1.3\pm0.15$. Observing a scaling with an exponent close to unity signals that the loss of global coherence is dominated by the identical spin-rotation effect, which scales proportional to the mean-field interaction strength $g$. In contrast, the rate of elastic collisions  scales as $\propto1/\ln(k_Fa_{2D})^2$, i.e., $\beta=2$ (see dashed line in Fig. 4c). This suggest the following decoherence mechanism in a spin-1/2 Fermi gas: after building up in the magnetic field gradient, the spin wave decays when neighbouring atoms have acquired a sufficiently large phase-angle $\Delta\phi$ between their spins such that in a spin-conserving collision both spins rotate out of the $S_x-S_y$-plane. Initially, this is a coherent process, which subsequently dephases due to the coupling of the spin-rotation to the momentum of the atoms\cite{Fuchs2002}.  This picture contrasts to the previously conjectured mechanism of the decoherence of a spin mixture resulting from a magnetic field curvature\cite{Gupta2003,Du2008}, which has been probed, for example, using the Ramsey technique\cite{Gupta2003}. Our spin-echo measurements show that the dephasing time measured by the Ramsey technique underestimates than the actual spin decoherence time in the weakly interacting regime by orders of magnitude, affecting the question of the equilibration of the Fermi gas.

We thank E. Altman, E. Demler, U. Ebling, A. Eckardt, C. Kollath, M. Lewenstein, A. Recati, W. Zwerger for discussions and B. Fr{\"o}hlich and M. Feld for early work on the experimental apparatus. The work has been supported by {EPSRC} (EP/J01494X/1, EP/K003615/1), the Leverhulme Trust (M. K.), the Royal Society, the Wolfson Foundation, and an Alexander-von-Humboldt Professorship.

Correspondence and requests for materials should be addressed to M. K. (email: mk673@cam.ac.uk) or M. K{\"o}hl~(email: michael.koehl@uni-bonn.de).

\section{methods}

\subsection{Preparation of the two-dimensional gases}
We prepare a quantum degenerate Fermi gas of $^{40}$K atoms in a one-dimensional optical lattice of wavelength $\lambda=1064$\,nm populating a stack of approximately 40 individual two-dimensional quantum gases\cite{Frohlich2011,Feld2011}. The radial confinement of the two-dimensional gases is harmonic with a trap frequency of $\omega_r=2\pi\times 127$\,Hz and the axial trap frequency is $\omega_z=2 \pi \times 75$\,kHz. Using a radio-frequency cleaning pulse, we ensure preparation of a spin-polarized gas in the $|F=9/2,m_F=-9/2\rangle\equiv \sket{\downarrow}$ ground state with $1.5\times 10^5$ atoms, corresponding to the Fermi energy of the spin-polarised gas  $E_F=12.3(2.0)\,$kHz at a temperature of $k_BT/E_F=0.24(3)$. The coupling strength of a two-dimensional Fermi gas is given by the parameter $g=-2 \pi \hbar^2/[m\ln(k_Fa_{2D})]$\cite{Bloom1975}. Here, $k_F$ is the Fermi wave vector for the initially spin-polarized sample, and $a_{2D}$ is the two-dimensional scattering length which is linked to the binding energy of the confinement-induced dimer by $E_B=\hbar^2/ma_{2D}^2$.

Our experiments on the spin dynamics start by ramping the magnetic field from $B=209.15\,$G to the desired value $B$ near the Feshbach resonance in $30\,$ms and wait for $200\,$ms. Subsequently we apply the spin-echo pulse sequence (see main text) of three radio-frequency pulses  with frequencies of $\sim50\,$MHz. The pulses have a  square envelope and duration of $t_{\pi/2}\sim46\,\mu$s for a $\pi/2$-pulse and $2\times t_{\pi/2}$ for a $\pi$-pulse. The pulses are separated by the same time $\tau$ which gives the total spin evolutions time $t=2\tau$.

\subsection{Ramsey measurements: Dephasing in the magnetic field gradient}
We assess the time scale of simple dephasing of the spin in the magnetic field gradient by performing Ramsey spectroscopy on the non-interacting gas using the following sequence: the initial $\pi/2$-pulse is followed by a second $\pi/2$-pulse with a variable time delay $\tau$. We observe a Ramsey coherence time of $\tau_{Ramsey}=(600\pm 30)\mu$s. Since the Ramsey time is more than an order of magnitude shorter than the inverse trap frequency, we neglect the motional contribution to the dephasing. Instead, we relate the Ramsey time to the dephasing time of the (quasi stationary) spins across the whole cloud, and we find from this the magnetic field gradient. We fit the data with $ P(t)=\frac{8 N J_2(\phi)}{\phi^2} $   Where, $ \phi = 2 \pi \Delta t = d\gamma B' R_F t$. $\Delta$ is the Zeeman shift due to the gradient at the edge of the cloud compared to the center. The fit function fits $ \phi = bt $, i.e., $ b = 2 \pi \Delta$. The fit results gives $ b = 8.1(1)\,$kHz and we know $ d \gamma = 152\,$kHz/G  at $209.15\,$G and $ R_F = 17.7\mu$m.

\end{document}